\documentclass[12pt]{article}%
\usepackage{ amsmath, graphics, rotating}%

\begin{document}
\begin{center}{\large\bf DOES THE COSMOLOGICAL CONSTANT PROBLEM EXIST?}\end{center}

\begin{center} {\large Felix M. Lev} \end{center}
\begin{center} {\it Artwork Conversion Software Inc.,
1201 Morningside Drive, Manhattan Beach, CA 90266, USA (Email: felixlev314@gmail.com}) \end{center}

\begin{flushleft} {\it Abstract:} \end{flushleft}
We first give simple arguments in favor of the "Zero Constants Party", i.e. that quantum theory should
not contain fundamental dimensionful constants at all. Then we argue that quantum theory should proceed not 
from a space-time background but from a Lie algebra, which is treated
as a symmetry algebra. With such a formulation of symmetry, the fact that $\Lambda\neq 0$ means not
that the space-time background  is curved (since the notion of the space-time background is not physical) but that 
the symmetry algebra is the de Sitter algebra rather than the Poincare one. In particular, there is 
no need to involve dark energy or other fields for explaining this fact. As a consequence, instead of the cosmological constant  
problem we have a problem why nowadays Poincare symmetry is so good approximate symmetry. 
This is rather a problem of cosmology but not fundamental quantum physics. 

\begin{flushleft} PACS: 11.30Cp, 11.30.Ly\end{flushleft}
\begin{flushleft} Keywords: cosmological constant problem, quantum theory, de Sitter invariance\end{flushleft}

\begin{sloppypar}
\section{Cosmological constant problem and units of measurement}
\end{sloppypar}

The literature on the cosmological constant problem is rather vast and numerous authors express very different opinions.
The history of this problem has become a folklore of theoretical physics. This history is described e.g. in Ref. 
\cite{EdS} and many other publications. 

We would like to begin our presentation with a discussion of another well known problem: how many independent 
dimensionful constants are needed for a complete description of nature? 
A paper \cite{Okun} represents a trialogue between three well known scientists:
M.J. Duff, L.B. Okun and G. Veneziano. The results of their discussions are summarized as
follows: {\it LBO develops the traditional approach with three constants, GV argues in favor of 
at most two (within superstring theory), while MJD advocates zero.} According to Weinberg 
\cite{W-units}, a possible definition of a fundamental constant might be such that it cannot
be calculated in the existing theory. We would like to give arguments in favor of the opinion
of the first author in Ref. \cite{Okun}. 

Consider a measurement of a component of angular momentum. The result depends on the system of units. 
As shown in quantum theory, in units $\hbar/2=1$ the result is given by an integer $0, \pm 1, \pm 2,...$. But we can
reverse the order of units and say that in units where the momentum is an integer $l$, its value in 
$kg\cdot m^2/sec$ is $(1.05457162\cdot 10^{-34}\cdot l/2)kg\cdot  m^2/sec$. Which of those two values
has more physical significance? In units where the angular momentum components are integers, the commutation
relations between the components are 
$$[M_x,M_y]=2iM_z\quad [M_z,M_x]=2iM_y\quad [M_y,M_z]=2iM_x$$
and they do not depend on any parameters. Then the meaning of $l$ is clear: it shows how big the angular momentum is in 
comparison with the minimum nonzero value 1. At the same time, the measurement of the angular momentum in units 
$kg\cdot m^2/sec$ reflects only a historic fact that at macroscopic
conditions on the Earth in the period between the 18th and 21st
centuries people measured the angular momentum in such units.

The fact that quantum theory can be written without the quantity $\hbar$ at all is usually treated as a choice
of units where $\hbar/2=1$ (or $\hbar=1$). We believe that a better interpretation of this fact is simply that quantum theory tells us
that physical results for measurements of the components of angular momentum should be given in integers.
Then the question why $\hbar$ is as it is, is not a matter of quantum physics 
since the answer is: because we want to measure components of angular momentum in $kg\cdot m^2/sec$. 

Our next example is the measurement of velocity $v$. The fact that any relativistic theory can be written without involving
$c$ is usually described as a choice of units where $c=1$. Then the quantity $v$ can take only values
in the range [0,1]. However, we can again reverse the order of units and say that relativistic theory tells us 
that results for measurements of velocity should be given by values in [0,1]. Then the question why $c$ is as it is, 
is again not a matter of physics since the answer is: because we want to measure velocity in $m/sec$. 

One might pose a question whether or not the values of $\hbar$ and $c$ may change with time. As far as $\hbar$ is concerned,
this is a question that if the angular momentum equals one then its value in $kg\cdot  m^2/sec$ will always be 
$1.05457162\cdot 10^{-34}/2$ or not. It is obvious that this is not a problem of fundamental physics but a problem how
the units $(kg,m,sec)$ are defined. In other words, this is a problem of metrology and cosmology. At the same time, the value
of $c$ will always be the same since the modern {\it definition} of meter is the length which light passes during $(1/(3\cdot 10^8))sec$.  

It is often believed that the most fundamental constants of nature are $\hbar$, $c$ and the gravitational constant $G$. The units
where $\hbar=c=G=1$ are called Planck units. Another well known notion is the $c\hbar G$ cube of physical theories. The meaning is
that any relativistic theory should contains $c$, any quantum theory should contains $\hbar$ and any gravitational theory 
should contain $G$. However, the above
remarks indicates that the meaning should be the opposite. In particular, relativistic theory {\it should not} contain $c$ and 
quantum theory {\it should not contain} $\hbar$. The problem of treating $G$ will be discussed below. 

A standard phrase that relativistic theory becomes nonrelativistic one
when $c\to\infty$ should be understood such that if relativistic theory is rewritten in conventional (but not physical!) units
then $c$ will appear and one can take the limit $c\to\infty$. A more physical descripion of the transition is that all the 
velocities in question are much less than unity. We will see below that those definitions are not equivalent.
Analogously, a more physical description of the transition from quantum to classical theory should be that all angular momenta
in question are very large rather than $\hbar\to 0$. 

Consider now what happens if we assume that de Sitter symmetry is fundamental. 
For definiteness we will discuss the de Sitter (dS) SO(1,4) symmetry and the same
considerations can be applied to the anti de Sitter (AdS) symmetry SO(2,3). The dS space-time is a four-dimensional
manifold in the five-dimensional space defined by 
\begin{equation}
x_1^2+x_2^2+x_3^2+x_4^2-x_0^2=R^2
\label{1}
\end{equation} 
In the formal limit $R\to\infty$ the action of the dS group in a large vicinity of the North or South pole of this manifold 
(i.e. when $x_4=\pm R$) becomes the action of the Poincare group on Minkowski space. In the literature, instead of $R$, 
the cosmological constant $\Lambda=3/R^2$ is often used. Then $\Lambda>0$ in the dS case and $\Lambda<0$ in the AdS one. 
Note that the dS space can be parametrized without using the quantity $R$ at all if instead of $x_a$ ($a=0,1,2,3,4$) we define
dimensionless variables $\xi_a=x_a/R$. It is also clear that  the elements of the SO(1,4) group
do not depend on $R$ since they are products of conventional and hyperbolic rotations. So the dimensionful value of $R$ 
appears only if one wishes to measure coordinates on the dS space in terms of coordinates of the flat five-dimensional space
where the dS space is embedded in. This requirement does not have a fundamental physical meaning. Therefore the value of $R$
defines only the scale factor for measuring coordinates in the dS space. By analogy with $c$ and $\hbar$,
the question  why $R$ is as it is, is not a matter of quantum theory since the answer is: because we want to measure distances in meters.

If one assumes that space-time background is fundamental then in the spirit of General Relativity (GR) it is natural to think that
the empty space-time is flat, i.e. that $\Lambda=0$ and this was the subject of the well-known 
dispute between Einstein and de Sitter. However, in view of the recent astronomical data,
it is now accepted that $\Lambda\neq 0$ and, although it is very small, it is positive rather than negative.

If we accept the parametrization of the dS space as in Eq. (\ref{1}) then the metric tensor on the dS space is obviously
\begin{equation}
g_{\mu\nu}=\eta_{\mu\nu}-x_{\mu}x_{\nu}/(R^2+x_{\rho}x^{\rho})
\label{2}
\end{equation}
where $\mu,\nu,\rho = 0,1,2,3$, $\eta_{\mu\nu}$ is the diagonal tensor with the components 
$\eta_{00}=-\eta_{11}=-\eta_{22}=-\eta_{33}=1$ and a summation over repeated indices is assumed. It is easy to calculate the
Christoffel symbols in the approximation where all the components of the vector $x$ are much less than $R$:
$\Gamma_{\mu,\nu\rho}=-x_{\mu}\eta_{\nu\rho}/R^2$.
Then a direct calculation shows that in the nonrelativistic approximation the equation of motion for a single particle is
\begin{equation}
{\bf a}={\bf r}c^2/R^2
\label{3}
\end{equation}
where ${\bf a}$ and ${\bf r}$ are the acceleration and the radius vector of the particle, respectively.

The fact that even a single particle in the Universe has a nonzero acceleration might be treated as contradicting the law of 
inertia but this law has been postulated only for Galilei or Poincare symmetries and we have ${\bf a}=0$ in the 
limit $R\to\infty$. A more serious problem arises
if GR is applied for describing a free particle in the dS world. According to GR, any system moving with an acceleration 
necessarily loses energy for
emitting gravitational waves. According to the Einstein quadrupole formula, the loss of the energy is given by  
$-dE/dt=(G/45c^5)(d^3D_{ik}/dt^3)^2$ where $G$ is the gravitational constant, $D_{ik}$ is the quadrupole moment and $i,k=1,2,3$. 
For a single particle moving
along the $x$ axis, the only nonzero element of the quadrupole moment is $D_{xx}=2mx^2$ where $m$ is the particle mass. 
Therefore, as follows from Eq. (\ref{3}), $-dE/dt=4Gx^2v^2/45c^3R^2$ where $v$ is the particle velocity. We see that the loss of 
energy depends on the choice of the origin in the coordinate space and one might think that this result is unphysical.

In the literature there are several different opinions on such a possibility. One might say that in the given case it is not 
legitimate to apply GR since the constant $G$ characterizes interaction between
different particles and cannot be used if only one particle exists in the world. Moreover, although GR has been confirmed in
several experiments in Solar system, it is not clear whether it can be extrapolated to cosmological distances.
More popular explanations are based on the assumption that the empty dS space cannot be literally empty. If the Einstein 
equations are written in the form
$G_{\mu\nu}+\Lambda g_{\mu\nu}=(8\pi G/c^4)T_{\mu\nu}$ where $T_{\mu\nu}$ is the stress-energy tensor of matter then the
case of empty space is often treated as a vacuum state of the field with the stress-energy tensor $T^{vac}_{\mu\nu}$ such that
$(8\pi G/c^4)T^{vac}_{\mu\nu}=-\Lambda g_{\mu\nu}$. This field is often called dark energy. With such an approach one implicitly
returns to Einstein's point of view that a curved space cannot be empty. Then the fact that $\Lambda\neq 0$ is treated as a dark energy
on the flat background. In other words, this is an assumption that Poincare symmetry is fundamental while dS one is
emergent.

However, in this case a new serious problem arises. The corresponding quantum theory is not renormalizable and with reasonable 
cutoffs the quantity
$\Lambda$ in units $\hbar/2=c=1$ appears to be of order $1/l_{P}^2=1/G$ where $l_P$ is the Planck length. 
It is obvious that since in the above theory the
only dimensionful quantities in units $\hbar/2=c=1$ are $G$ and $\Lambda$, and the theory does not have other parameters, the
result that $G\Lambda$ is of order unity seems to be natural. However, this value of $\Lambda$ is at least by 120 orders of magnitude
greater than the experimental one. Numerous efforts to solve this cosmological constant problem have not been successful so far although
many explanations have been proposed. In addition, many physicists argue that in the spirit
of GR, the theory should not depend on the choice of the background space-time (so called a principle of background independence)
and there should not be a situation when the flat background is preferable.

\section{Cosmological constant problem in quantum theory}

Consider now the dS symmetry from the point of view of quantum theory. In this theory any physical quantity can be discussed 
only in conjunction with the operator defining this quantity. For example, in standard quantum mechanics  
the quantity $t$ is a parameter, which has the meaning of time only in the
classical limit since there is no operator corresponding to this quantity. The problem of how time should be defined
on quantum level is very difficult and is discussed in a vast literature. It has been also well known  
since the 1930's \cite{NW} that, when quantum mechanics is combined with 
relativity, there is no operator satisfying all the properties of the spatial position operator. In other 
words, the coordinates cannot be exactly measured even in situations when exact measurements are allowed by the
nonrelativistic uncertainty principle. In the introductory section of the well-known textbook \cite{BLP}
simple arguments are given that for a particle with mass $m$, the coordinates cannot be measured with the accuracy 
better than the Compton wave length ${\hbar}/mc$. Hence, the exact measurement is possible only
either in the nonrelativistic limit (when $c\to\infty$) or classical limit (when ${\hbar}\to 0)$.

Let us proceed from the following principle: definition of a physical quantity is a description how this quantity 
should be measured. From this point of view, one can discuss if {\it coordinates of particles} 
can be measured with a sufficient accuracy, while the notion
of space-time background, regardless of whether it is flat or curved, does not have a physical meaning. Indeed, this notion 
implies that space-time coordinates are meaningful even if they refer not to real
particles but to points of a manifold which exists only in our imagination. However, such coordinates are not measurable. 
To avoid this problem one might try to treat the space-time background as a reference frame.
Note that even in GR, which is a pure classical (i.e. non-quantum) theory, the meaning of reference frame is not 
clear. In standard textbooks (see e.g. Ref. \cite{LL}) the reference frame in GR is defined as a collection 
of weightless bodies, each of which is characterized by three numbers (coordinates) and is supplied by a clock. 
Such a notion (which resembles ether) is not physical even on classical level and for sure it is meaningless 
on quantum level. There is no doubt that GR is a great achievement of theoretical physics and has achieved 
great successes in describing experimental data. At the same time, it is based on the notions of space-time background or 
reference frame, which do not have a clear physical meaning. Therefore it is unrealistic to expect that successful 
quantum theory of gravity will be based on quantization of GR. The results of GR should follow from quantum theory of 
gravity only in situations when space-time coordinates of real bodies is a good approximation while in general the 
formulation of quantum theory might not involve space-time at all.

In particular, the quantity $x$ in the Lagrangian density $L(x)$ is not measurable. Note that the Lagrangian density 
is only an auxiliary tool for deriving equations of motion in classical theory and constructing Hilbert spaces and 
operators in quantum theory. After this construction has been done, one can safely forget about  
background space-time coordinates and Lagrangian. So Lagrangian can be at best treated
as a hint for constructing a reasonable theory since a fundamental approach should not proceed from
notions, which have no meaning. As stated in Ref. \cite{BLP},
local quantum fields and Lagrangians are rudimentary notion, which will disappear in the ultimate quantum theory.
Those ideas have much in common with the Heisenberg S-matrix program
and were rather popular till the beginning of the 1970's. Although no one questioned those ideas, they
are now almost forgotten in view of successes of gauge theories.

If we accept that quantum theory should not proceed from space-time background, a problem arises how symmetry 
should be defined on quantum level. In the spirit of Dirac's
paper \cite{Dir}, we postulate that on quantum level a symmetry means that a system is described by a set 
of operators, which satisfy certain commutation relations. 
We believe that for understanding this Dirac's idea the following example might be useful.
If we define how the energy should be measured (e.g. the energy of bound states, kinetic energy etc.),
we have a full knowledge about the Hamiltonian of our system. In particular, we know how the Hamiltonian
should commute with other operators. In standard theory the Hamiltonian is also interpreted as an
operator responsible for evolution in time, which is considered as a classical macroscopic parameter. 
In situations when this parameter is a good approximate parameter, macroscopic transformations
from the symmetry group corresponding to the evolution in time have a physical meaning. However, there is no guarantee that 
such transformations always have a physical meaning (e.g. at the very early stage of the Universe). 
In general, according to principles of quantum theory, selfadjoint operators in Hilbert spaces represent observables but
there is no requirement that parameters defining a family of unitary transformations generated by a selfadjoint operator
are eigenvalues of another selfadjoint operator. A well known example from standard quantum mechanics is that if $P_x$ is
the $x$ component of the momentum operator then the family of unitary transformations generated by $P_x$ is $exp(iP_xx/\hbar)$
where $x\in (-\infty,\infty)$ and such parameters can be identified with the spectrum of the position operator. At the same time, 
the family of unitary transformations generated by the Hamiltonian $H$ is $exp(-iHt/\hbar)$ where $t\in (-\infty,\infty)$
and those parameters cannot be identified with a spectrum of a selfadjoint operator on the Hilbert space of our system. 
In the relativistic case the parameters $x$ can be formally identified with the spectrum of the Newton-Wigner position
operator \cite{NW} but it is well known that this operator does not have all the required properties for the position
operator.

The {\it definition} of the dS symmetry on quantum level is that the operators $M^{ab}$ 
($a,b=0,1,2,3,4$, $M^{ab}=-M^{ba}$) 
describing the system under consideration satisfy the commutation relations {\it of the dS Lie algebra} so(1,4), i.e. 
\begin{equation}
[M^{ab},M^{cd}]=-2i (\eta^{ac}M^{bd}+\eta^{bd}M^{ac}-
\eta^{ad}M^{bc}-\eta^{bc}M^{ad})
\label{4}
\end{equation}
where $\eta^{ab}$ is the diagonal metric tensor such that
$\eta^{00}=-\eta^{11}=-\eta^{22}=-\eta^{33}=-\eta^{44}=1$.
These relations do not depend on any free parameters. One might say that this is a consequence of the choice of units
where $\hbar/2=c=1$. However, as noted above, any fundamental theory should not involve the quantities $\hbar$ and $c$.

With such a definition of symmetry on quantum level, the dS symmetry looks more natural than the Poincare symmetry.
In the dS case all the ten representation operators of the symmetry algebra are angular momenta while in the
Poincare case only six of them are angular momenta and the remaining four operators represent standard energy and
momentum. If we define the operators $P^{\mu}$ as $P^{\mu}=M^{4\mu}/R$ then in the formal limit when $R\to\infty$,
$M^{4\mu}\to\infty$ but the quantities $P^{\mu}$ are finite, the relations 
(\ref{4}) will become the commutation relations for representation operators of the Poincare algebra such that the 
dimensionful operators $P^{\mu}$ are the four-momentum operators. 

A theory based on the above definition of the dS symmetry on quantum level cannot involve quantities which are 
dimensionful in units $\hbar/2=c=1$. In particular, we inevitably come to conclusion that the dS space, 
the gravitational constant and the cosmological constant
cannot be fundamental. The latter appears only as a parameter replacing the dimensionless operators $M^{4\mu}$ by
the dimensionful operators $P^{\mu}$ which have the meaning of momentum operators only if $R$ is rather large. 
Therefore the cosmological constant problem does not arise at all
but instead we have a problem why nowadays Poincare symmetry is so good approximate symmetry. This is rather a problem of 
cosmology but not quantum physics. 

The next question is how elementary particles in quantum theory should be defined. A discussion of numerous controversial 
approaches can be found, for example in the recent paper \cite{Rovelli}.
Since we do not accept approaches based on the background space-time, we accept an approach where, by definition, elementary particles in 
the dS invariant theory are described
by irreducible representations (IRs) of the dS algebra by Hermitian operators. As shown in Refs. \cite{lev1a,lev3}, such representations 
can be explitly constructed by using well known results about unitary irreducible representations (UIRs) of the dS group. 
An excellent description of such UIRs for physicists can be found in a book by Mensky \cite{Mensky}. As shown in Ref. \cite{Mensky},
they can be implemented in the Hilbert space of functions $f({\bf v})$ defined on two Lorentz hyperboloids  
$v_0=\pm (1+{\bf v}^2)^{1/2}$ such that $\int |f({\bf v})|^2d^3{\bf v}/|v_0|<\infty$. 

In Refs. \cite{lev1a,lev3} we have described all the technical details
needed for computing the explicit form of the generators $M^{ab}$.
In the spinless case, the action  of the generators on functions defined on the upper hyperboloid is
\begin{eqnarray}
&&{\bf M}=2l({\bf v}),\quad {\bf N}==-2i v_0 \frac{\partial}{\partial {\bf v}}\quad 
M_{04}=m_{dS} v_0+2i v_0({\bf v}\frac{\partial}{\partial {\bf v}}+\frac{3}{2})\nonumber\\
&& {\bf B}=m_{dS} {\bf v}+2i [\frac{\partial}{\partial {\bf v}}+
{\bf v}({\bf v}\frac{\partial}{\partial {\bf v}})+\frac{3}{2}{\bf v}]
\label{5}
\end{eqnarray}
where ${\bf M}=\{M^{23},M^{31},M^{12}\}$,
${\bf N}=\{M^{01},M^{02},M^{03}\}$,
${\bf B}=\{M^{41},M^{42},M^{43}\}$, ${\bf l}({\bf v})=-i{\bf v}
\times \partial/\partial {\bf v}$ and $m_{dS}$ is the dS mass of the particle.
The expressions for ${\bf M}$ and ${\bf N}$ are the same as in the case of Poincare symmetry but the expressions
for $M_{04}$ and ${\bf B}$ are different.

Note that in deriving these expressions no approximations have been made and the results are exact. 
In particular, the dS space, the cosmological constant and 
the Riemannian geometry have not been involved at all. Nevertheless, the expressions for the representation operators
is all we need to have the maximum possible information in quantum theory.

We now define $E=M_{04}/R$, ${\bf P}={\bf B}/R$ and $m=m_{dS}/R$. Consider the nonrelativistic approximation when $|{\bf v}|\ll 1$.
If we wish to work with units where the dimension of velocity is $m/sec$, we should
replace ${\bf v}$ by ${\bf v}/c$. If ${\bf p}=m{\bf v}$ then it is clear from the expression for ${\bf B}$ that ${\bf p}$
becomes the real momentum ${\bf P}$ only in the limit $R\to\infty$. Now by analogy with nonrelativistic quantum mechanics, 
we {\it define} the position operator ${\bf r}$ as $2i\partial/\partial {\bf p}$ (since we formally accept units where 
$\hbar/2=1$ rather than $\hbar=1$). In classical approximation we can treat ${\bf p}$ and ${\bf r}$ as usual vectors
and neglect their commutators. Then the results for the classical nonrelativistic energy and momentum are
\begin{equation}
H=\frac{{\bf p}^2}{2m} + \frac{c{\bf p}{\bf r}}{R}\quad {\bf P}={\bf p}+\frac{mc{\bf r}}{R}
\label{6}
\end{equation} 
where $H=E-mc^2$ is the classical energy. As follows from these expressions, the classical Hamiltonian is 
\begin{equation}
H({\bf P},{\bf r})=\frac{{\bf P}^2}{2m}-\frac{mc^2{\bf r}^2}{2R^2}
\label{7}
\end{equation}

The last term in Eq. (\ref{7}) is the dS correction to the nonrelativistic Hamiltonian. It is interesting to note that
the nonrelativistic Hamiltonian depends on $c$ although it is usually believed that $c$ can be present only in
relativistic theory. This illustrates the fact mentioned in the beginning of the paper that the transition to
nonrelativistic theory understood as $|{\bf v}|\ll 1$ is more physical than that understood as $c\to\infty$. The presence of
$c$ in Eq. (\ref{7}) is a consequence of the fact that this expression is written in standard units. In nonrelativistic
theory $c$ is usually treated as a very large quantity. Nevertheless, the last term in Eq. (\ref{7}) is not large since
we assume that $R$ is very large.  

The result given by Eq. (\ref{3}) is now a consequence of the equations of motion for the Hamiltonian given by Eq. (\ref{7}).
In our approach this result has been obtained without using dS space and Riemannian geometry while  
the fact that $\Lambda\neq 0$ should be treated not such that the background space-time has a curvature (since the notion of the
background space-time is meaningless) but as an indication that the symmetry algebra is the dS algebra rather than the Poincare one.
{\it Therefore for explaining the fact that $\Lambda\neq 0$ there is no need to involve dark energy or any other quantum fields.}

In quantum theory the Fock space for a given quantum system is a tensor product of Hilbert spaces describing elementary
particles. In particular, a two-particle Hilbert space is a tensor product of the single-particle spaces. If the
particles do not interact, then, by definition, representation operators describing a two-particle representation are sums of 
the corresponding
single-particle operators. So in the dS invariant theory one can use the results for IRs and calculate the mass operator of the
free two-body system. The result of calculations \cite{lev1a,lev3} is that in the approximation when
the relative distance operator can be defined with a good accuracy, the additional term in the nonrelativistic mass operator
in comparison with the Poincare theory is $V_{dS}(r) = -m_{12}r^2/(2R^2)$ where now $r$ is the relative distance and 
$m_{12}=m_1m_2/(m_1+m_2)$ is the reduced mass. As a consequence, in quasiclassical approximation the relative 
acceleration is given by the same expression (\ref{3}) but now ${\bf a}$ is the relative acceleration and ${\bf r}$ is the relative
radius vector.    
 
The fact that two free particles have a relative acceleration is well known for cosmologists who 
consider the dS symmetry on classical level. This effect is called the dS antigravity. 
The term antigravity in this context means that the particles repulse rather than attract each other. In the case of the 
dS antigravity the relative acceleration of two free particles is proportional (not inversely proportional!) to the distance 
between them. This classical result (which in our approach has been obtained without involving dS space and Riemannian geometry)
is a special case of the dS symmetry on quantum level when quasiclassical approximation works with a good accuracy. 
At the same time, our discussion shows that in dS invariant theory, free particles will not emit gravitational waves since 
the existence of relative acceleration has nothing to do with dark energy or other fields. This fact is
also related to the above remark that the dS symmetry on quantum level excludes $G$ from being a fundamental quantity. 
It is well known that at present the phenomenon of gravity has been observed only at macroscopic conditions.
Also there exists a vast literature discussing a possibility that gravity is not fundamental but emergent.

A possible approach for seeking new theories might be based on finding new symmetries such that known symmetries are
special cases of the new ones when a contraction parameter goes to zero or infinity (see e.g. the famous paper \cite{Dyson} 
entitled "Missed Opportunities"). For example, classical theory is a special case of quantum one when $\hbar\to 0$ and
nonrelativistic theory is a special case of relativistic one when $c\to\infty$. From this point of view, de Sitter symmetry
is "better" than Poincare one since the latter is a special case of the former when $R\to\infty$. A question arises whether
there exists a ten-dimensional algebra, which is more general than the dS one, i.e. the dS algebra is a special case of this
hypothetical new algebra when some parameter goes to zero or infinity. As noted in Ref. \cite{Dyson}, the answer is "no" since
the dS algebra is semisimple. So one might think that the only way to extend the de Sitter symmetries is to consider higher 
dimensions and this is in the spirit of modern trend.

However, if we consider a quantum theory not over complex numbers but over a Galois field of characteristic $p$ then
standard dS symmetry can be extended as follows. We require that the operators $M^{ab}$
satisfy the same commutation relations as above but those operators are considered in spaces over a Galois field.
Such operators implicitly depend on $p$ but they still do not depend on $R$. This approach, which we
call quantum theory over a Galois field (GFQT), has been discussed in details in Refs. \cite{lev3,lev2}.
GFQT is a more general theory than the standard one since the latter is a special case
of the former when $p\to\infty$. In the approximation when $p$ is very large, GFQT can reproduce all the standard
results of quantum theory. At the same time, GFQT is well defined mathematically since it does not contain
infinities. Note that while in standard theory the dS and AdS algebras are
"better" than the Poincare algebra from aesthetic considerations (see the discussion above), in GFQT there is no choice
since Poincare algebra over a Galois field is unphysical (see the discussion in Refs. \cite{lev3,lev2}). 

In view of the above discussion, it seems natural to express all dimensionful quantities in terms of $(c,\hbar,R)$ rather
than $(c,\hbar,G)$ since the former is a set of parameters characterizing transitions from higher symmetries to lower ones.
Then a reasonable question is why the quantity $G$ is so small. Indeed, in units $\hbar/2=c=1$, $G$ has the dimension $length^2$
and so one might expect that it should be of order $R^2=3/\Lambda$. So again the disagreement is more that 120 orders of magnitude
and one might call this the gravitational constant problem rather than the cosmological constant problem. As noted above, 
in standard theory a reasonable possibility is that $G\Lambda$ is of order unity. However, in GFQT we have a
parameter $p$. In Ref. \cite{essay} we have described our hypothesis that $G$ contains a factor $1/lnp$ and that is why it is so small. 

\section{Conclusion}

The main achievements of modern theory have been obtained in the aproach proceeding from space-time background.
In quantum theory this approach is not based on a solid mathematical basis and, as a consequence, the problem of infinities
arises. While in QED and other renormalizable theories this problem can be somehow circumvented, in quantum gravity this 
is not possible even in the lowest orders of perturbation theory. Mathematical problems of quantum theory are discussed in 
a wide literature. For example, in the
well known textbook \cite{Bogolubov} it is explained in details that interacting quantized fields can be treated 
only as operatorial distributions and hence their product at the same point is not well defined.
One of ideas of the string theory is that if a point (a zero-dimensional object) is replaced by a string
(a one-dimensional object) then there is hope that infinities will be less singular. 

For majority of physicists the fact that GR and quantum theory describe many experimental data with an unprecedented 
accuracy is much more important than a lack of mathematical rigor and that the notion of space-time
background is not physical. For this reason physicists do not wish to abandon this notion.
As one of the consequences, the cosmological constant problem arises and it is now believed that dark energy 
accounts for more than 70\% of the total energy of the Universe.
There exists a vast literature where different authors propose different approaches 
and some of the authors claim that they have found the solution of the problem. Meanwhile the above discussion clearly
demonstrates that the cosmological constant problem (which is often called the dark energy problem) 
is a purely artificial problem arising as a result of using
the notion of space-time background while this notion is not physical.

{\it Acknowledgements: } L.A. Kondratyuk and S.N. Sokolov paid my attention to Dirac's paper \cite{Dir}. 
They explained that the theory should not necessarily be based on a local Lagrangian, and symmetry on 
quantum level means that proper commutation relations are satisfied. 
E.G. Mirmovich proposed an idea that only angular momenta are fundamental physical quantities \cite{Mirmovich}.
I am also greatful to Volodya Netchitailo for discussions about fundamental physical constants.

\end{document}